\documentclass[notitlepage,showpacs,preprintnumbers,amsmath,amssymb,showkeys,aps,prl,10pt]{revtex4-1}

\usepackage{graphicx}
\usepackage{hyperref}
\usepackage{url}

\begin{document}

\title{Enantiomer-Specific State Transfer of Chiral Molecules}
\author{Sandra Eibenberger, John Doyle, and David Patterson}
\email{dave@cua.harvard.edu}
\email{sandra@cua.harvard.edu}

\affiliation{Harvard University, Department of Physics, 17 Oxford Street, Cambridge, MA 02138 USA}

\begin{abstract}
State-selective enantiomeric excess is realized using microwave-driven coherent population transfer. The method selectively promotes either $R$ or $S$ molecules to a higher rotational state by phase-controlled microwave pulses that drive electric-dipole allowed rotational transitions. We demonstrate the enantiomer-specific state transfer method using enantiopure samples of 1,2-propanediol. This method of state-specific enantiomeric enrichment can be applied to a large class of asymmetric, chiral molecules that can be vaporized and cooled to the point where rotationally resolved spectroscopy is possible, including molecules that rapidly racemize. The rapid chiral switching demonstrated here allows for new approaches in high-precision spectroscopic searches for parity violation in chiral molecules.

\end{abstract}


\maketitle

Chirality plays a major role in many biological processes and chemical reactions. A large number of pharmaceutically and biologically relevant molecules are chiral and exist in one of two mirror images (enantiomers). Although different enantiomers share many physical and chemical properties, their handedness often determines their functionality. Despite the high abundance of chiral species in nature, detecting and quantifying the handedness of a sample remains challenging. The origin of the observed biomolecular homochirality on Earth is unknown. A proposed explanation of this is parity violation (PV) effects in chiral molecules, that are predicted to cause tiny energy differences between enantiomers~\cite{Quack2008,Schwerdtfeger2010,Darquie2010}. These PV effects have yet to be experimentally measured.

There are a number of established spectroscopic techniques for chiral molecules that measure enantiomeric excess, which is a measure for the purity of a chiral sample, defining the excess of one enantiomer over the other in a mixture. These techniques include circular dichroism, vibrational circular dichroism, Raman optical activity~\cite{He2011,Stephens1985,Fanood2015}, and, most recently, enantiomer-specific microwave spectroscopy (EMS)~\cite{Patterson2013,Patterson2013a}. EMS is inherently mixture compatible and can yield a large signal, although it requires the sample to be vaporized and cooled. A natural extension of measuring enantiomeric excess (EE) is the separation of enantiomers from a sample of mixed chirality - an important and challenging task. The most widely used separation techniques are chromatography and capillary electrophoresis~\cite{Gubitz2000,Gubitz2001,Chankvetadze2007,Cavazzini2011,Beesley1999}, both of which require specific development and optimization of intricate processes for different molecular species. All current demonstrated enantiomer separation methods rely on \emph{chemical} mechanisms and are based on enantiomer-specific interactions with auxiliary substances, for example in a solution or in an enantioselective chromatographic column. The required time for chiral separation in these techniques is significantly longer than seconds, making chiral separation of rapidly racemizing compounds out of reach.

Here we report enhancement of enantiomeric excess in a specific rotational state using an approach solely based on microwave three-wave mixing~\cite{Patterson2013,Patterson2013a}. The enhancement of either the $S$ or $R$ enantiomer population is shown to be controllable by the phase of a single microwave field. Our method is an extension of EMS, where enantiomers are driven with two orthogonally polarized, oscillating, resonant electric fields, and the coherent radiation of the free induction decay (FID) is detected in a third, mutually orthogonal polarization. The emitted radiation has an intensity that is proportional to the product of the molecule's electric dipole moments $\vec{\mu}_a$, $\vec{\mu}_b$, $\vec{\mu}_c$. The triple product $\vec{\mu}_a \cdot (\vec{\mu}_b \times \vec{\mu}_c)$ is independent on the choice of the inertia principle axes $a, b, c$, but it changes sign for different enantiomers~\cite{Grabow2013}.
The chirality is therefore imprinted onto the phase of the resulting radiation, which is opposite for different enantiomers. This technique has been demonstrated on a variety of molecules, including 1,3-butanediol~\cite{Patterson2013}, carvone~\cite{Shubert2014}, 1,2-propanediol~\cite{Patterson2013a}, and menthone~\cite{Patterson2014}.

\begin{figure}[ht]
\includegraphics[width=0.5\columnwidth]{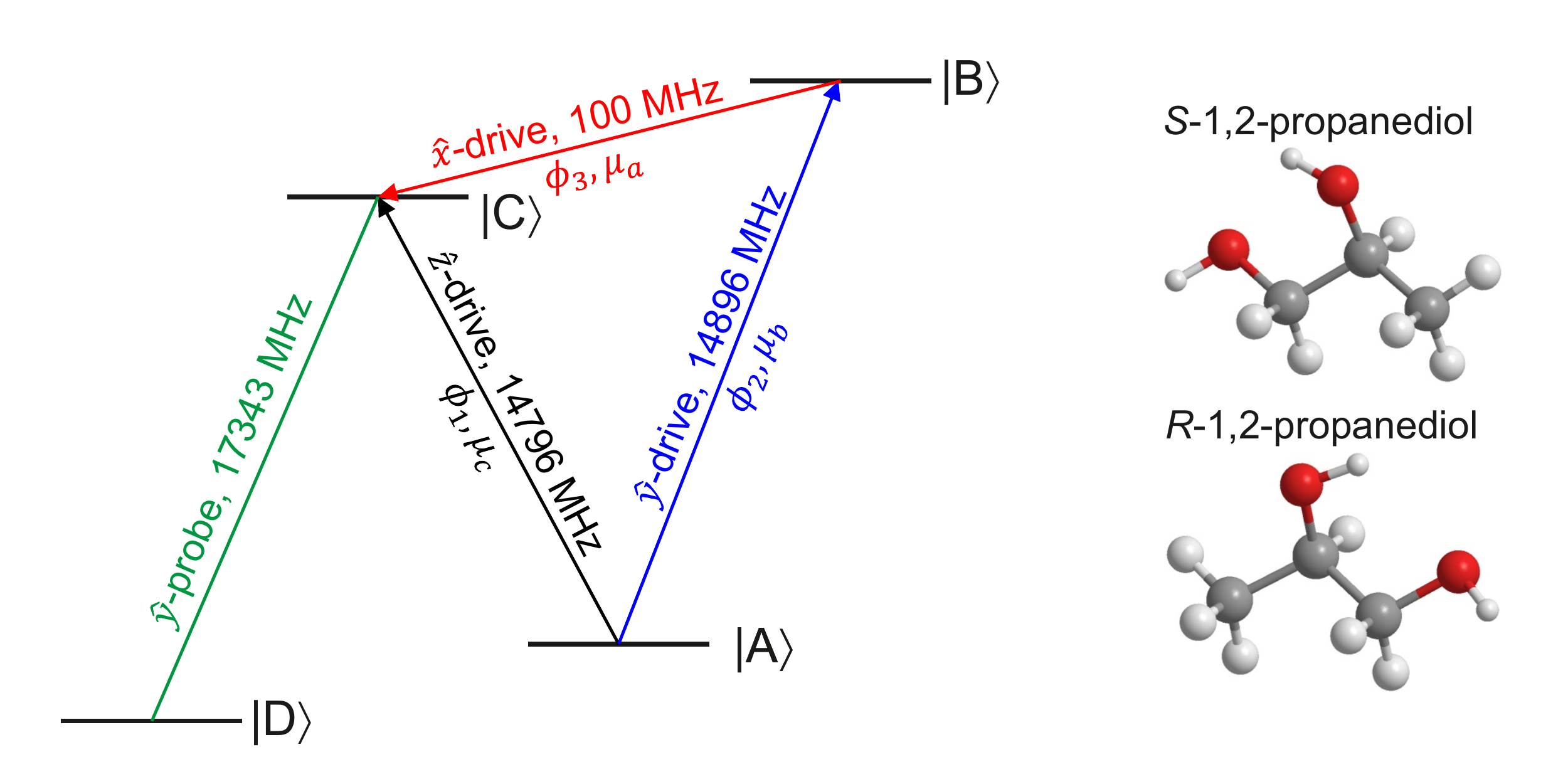} 
\caption{Left: Level diagram of the rotational states of 1,2-propanediol used for enantiomer-specific enrichment of a chosen rotational state. The relevant states are labeled $|A\rangle$, $|B\rangle$, $|C\rangle$, and $|D\rangle$. In our method, state $|C\rangle$ is populated on two paths - directly from $|A\rangle \rightarrow |C\rangle$ and indirectly $|A\rangle \rightarrow |B\rangle \rightarrow |C\rangle$. Depending on the phases $\phi_1$, $\phi_2$, and $\phi_3$ of the driving transitions, $|C\rangle$ is populated or depopulated for opposite enantiomers. The final population in the target state $|C\rangle$ is probed by using an additional, radiating transition $|C\rangle \rightarrow |D\rangle$. Right: Molecular structure of the lowest energy conformer of $S$ and $R$ 1,2-propanediol~\cite{Lovas2009}. \label{fig:leveldiagram}
}
\end{figure}

In our method of enantiomer-specific state transfer, enantiomers are driven with three, orthogonally polarized, resonant, phase-controlled electric fields.
Species-, conformer-, state- and enantiomer- specific population transfer between two rotational states of a chiral molecule, 1,2-propanediol, is achieved. The key feature of this method is enantiomer-specific population or depopulation of a selected rotational state. The relevant states are referred to as $|A\rangle$, $|B\rangle$, $|C\rangle$ in the level diagram of 1,2-propanediol (figure~\ref{fig:leveldiagram}). The molecules are subjected to microwave pulses that transfer state population by two distinct paths: directly from $|A\rangle \rightarrow |C\rangle$ and indirectly from $|A\rangle \rightarrow |B\rangle \rightarrow |C\rangle$. The amplitude of state $|C\rangle$ is comprised of contributions from each path, which interfere constructively or destructively depending on enantiomer. This results in opposite, state-specific enantiomeric enrichment in states $|A\rangle$ and $|C\rangle$. The final population in $|A\rangle$ or $|C\rangle$ is probed by driving additional transitions from these states, as in traditional Fourier transform microwave spectroscopy (FTMW).

In order to achieve significant enantiomeric enrichment in a specific rotational state, molecules must initially be sufficiently cold such that there is a substantial population difference in the relevant rotational states. For the initial population in $|C\rangle$ to be substantially lower than the initial population in state $|A\rangle$, a gas phase sample with a rotational and translational temperature of a few Kelvin is required. Such samples are routinely prepared in supersonic jet expansions or, as used here, in a cryogenic buffer gas~\cite{Patterson2012}. Cryogenic buffer gas cooling provides a continuous, high density sample of cold molecules and is particularly well suited for spectroscopy of complex mixtures and radicals. In addition, its continuous nature allows for high repetition rates (typically $\sim 60\,\mathrm{kHz}$) of the experimental cycle. This is approximately a factor of $10^{4}$ higher than in typical experiments using supersonic jet expansion. This, and the utilization of cryogenic low noise amplifiers, results in substantially increased sensitivity.

\begin{figure}[ht]
\includegraphics[width=\columnwidth]{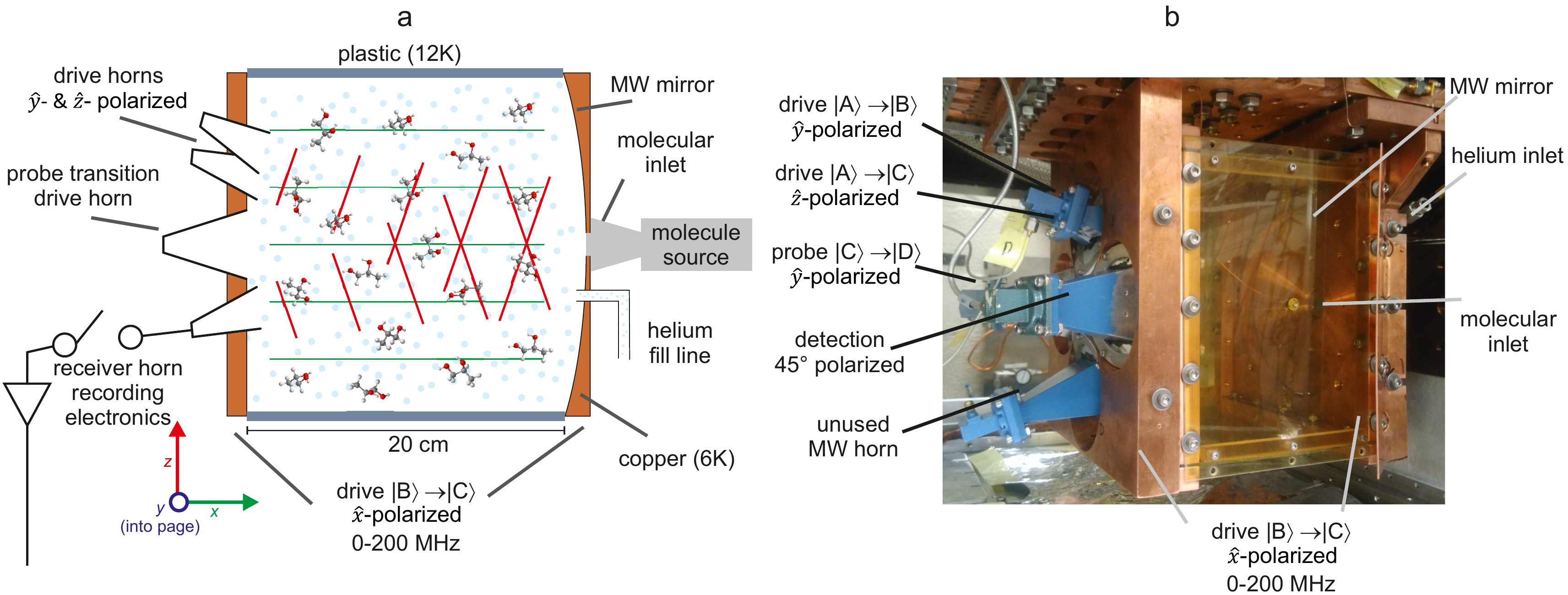} 
\caption{a) Sketch (not to scale) and b) picture of the experimental setup. 1,2-propanediol molecules are evaporated in front of the cell entrance and enter the cell through a hole in the microwave mirror. The cryogenic buffer gas cell is cooled by a closed-cycle pulse tube refrigerator ($4\,\mathrm{K}$). The hot molecules enter the cell and rapidly cool to approximately $10\,\mathrm{K}$. Microwave pulses of $\hat{y}$ and $\hat{z}$ polarization are introduced via microwave horns. An electric field of $\hat{x}$ polarization is produced between the microwave mirror and the opposite face of the buffer gas cell. A third microwave horn drives a transition from $|C\rangle\rightarrow|D\rangle$ for probing the population in state $|C\rangle$. The resulting FID is collected with an additional microwave horn, amplified with a cryogenic low noise amplifier, mixed down and sent to recording electronics. \label{fig:setup}
}
\end{figure}

The apparatus used here is similar to the one used in \cite{Patterson2013a}, with recent improvements to allow the introduction of microwave fields of arbitrary polarization. A sketch and a picture of the apparatus are shown in figure~\ref{fig:setup}. At the heart of the experiment is a cell $(\mathrm{20\,cm}\times\mathrm{20\,cm}\times\mathrm{20\,cm})$, which is cooled to about $T=10\,\mathrm{K}$. Via a gas fill line, helium is continuously introduced into the cell and the atoms thermalize with the cold cell walls. Warm, gas-phase molecules are produced by heating a liquid sample placed $\sim 2\,\mathrm{cm}$ in front of the cell entrance aperture. Within the cell, warm molecules and cold helium atoms mix, and the molecules rapidly cool to the temperature of the cell walls via helium-molecule collisions. Under typical conditions, we have $T_{\mathrm{rot}} \approx T \approx 10\, \mathrm{K}$, a helium density of $n_{\mathrm{He}} \approx 10^{14}\,\mathrm{cm}^{-3}$, and a molecule density of $n_{\mathrm{mol}} \approx 10^{11}\,\mathrm{cm}^{-3}$. Molecules collide with helium atoms approximately every $\tau_c \approx 6\,\mathrm{\mu s}$, randomizing the orientation of the molecules and thus destroying rotational state coherence on a time scale $\sim \tau_c$. Thus, $\tau_c^{-1}$ sets an approximate upper bound on the accessible repetition rate of the experiment. Two pairs of microwave horns produce microwave fields with $\hat{y}$ and $\hat{z}$ polarizations and frequencies of $12\,\mathrm{GHz} < f_{\mathrm{y,z}} < 18\,\mathrm{GHz}$. The cell wall containing the entrance aperture is electrically insulated from the opposite wall and can be driven directly to produce $\hat{x}-$polarized electric fields with $0\,\mathrm{MHz} \leq f_{\mathrm{x}} < 200\,\mathrm{MHz}$. 

Enantiomer-specific rotational state transfer requires a suitable set of levels appropriate for microwave three-wave mixing. A vast number of polar, asymmetric, chiral molecules ($C_1$ point group, no plane or axis of symmetry) possess one or more suitable sets of rotational states. A typical set of usable states is shown in figure~\ref{fig:leveldiagram}, where we identify an eligible set of rotational states of our example species, 1,2-propanediol. For clarity, the states are marked $|A\rangle$, $|B\rangle$, $|C\rangle$, $|D\rangle$, corresponding to $|A\rangle = |211\rangle$, $|B\rangle = |220\rangle$, $|C\rangle = |221\rangle$, and $|D\rangle = |212\rangle$, in the $|JK_aK_c\rangle$ notation.

The transitions are driven with a sequence of microwave pulses having $\hat{x}$, $\hat{y}$, and $\hat{z}$ polarizations. 
In the weak pulse limit, the pulse sequence can be understood as follows. Molecules begin preferentially populating the lowest energy state of this group of states $|A\rangle$. A weak pulse at frequency $f_{AC}$ and phase $\phi_1$ puts the molecules in a superposition $|A\rangle$ + $\alpha e^{i\phi_1}\mu_c|C\rangle$, where $\alpha$ is a positive, real-valued constant that is proportional to the amplitude and length of the pulse. Further pulses at frequency $f_{AB}$ with phase $\phi_2$ and frequency $f_{BC}$ with phase $\phi_3$ put the molecules in the superposition state $|A\rangle + \alpha e^{i\phi_1}\mu_c|C\rangle + \beta e^{i\phi_2}\mu_b|B\rangle + \beta\gamma\ e^{i\phi_2} e^{i\phi_3}\mu_b\mu_a|C\rangle + \alpha\gamma e^{i\phi_1} e^{i\phi_3}\mu_a\mu_c|B\rangle$. Here, $\beta$ and $\gamma$ are positive, real-valued constants proportional to the pulse amplitudes and durations, and $\mu_a$, $\mu_b$, and $\mu_c$ are the molecules' electric dipole moment components. The probability $P_{|C\rangle}$ for the molecule to be left in state $|C\rangle$ is $P_{|C\rangle} \propto |\alpha e^{i\phi_1}\mu_c + \beta\gamma\ e^{i(\phi_2+\phi_3)}\mu_a\mu_b|^2$.

We experimentally choose pulse lengths and intensities such that the two paths have equal magnitude, i.e. $\alpha=|\mu_a \mu_b||\mu_c|^{-1} \beta\gamma$. This yields
\begin{equation}\label{eq1}
P_{|C\rangle} \propto ||\mu_a \mu_b||\mu_c|^{-1}| \beta\gamma e^{i\phi_1} \mu_c + \beta\gamma\ e^{i(\phi_2+\phi_3)}\mu_c||^2.
\end{equation}

We can choose axes such that $\mu_a > 0$ and $\mu_b > 0$, thus forcing the sign of $\mu_c$ to change with enantiomer. Therefore, equation~\ref{eq1} becomes
\begin{equation}\label{eq2}
P_{|C\rangle} \propto |e^{i(\phi_2 + \phi_3)} \pm e^{i \phi_1}|^2 = 2[1 \pm \cos(\phi_2 + \phi_3 - \phi_1)],
\end{equation}

with the sign of $\pm$ depending on the enantiomer. The population in the target state $|C\rangle$ is probed by driving the transition $|C\rangle \rightarrow |D\rangle$  with a frequency $f_{CD}$, resulting in a FID. It should be noted that the phase $\phi_4$ of this probe pulse is not crucial to the technique and has no relation to any of the three driving phases $\phi_1$, $\phi_2$, and $\phi_3$. In fact, this probe could be replaced by any state-selective measurement, including incoherent methods, such as laser-induced fluorescence (LIF) or resonance-enhanced multiphoton ionization (REMPI) detection. Future use of these methods could increase the sensitivity of the readout by orders of magnitude. 

\begin{figure}[ht]
\includegraphics[width=0.5\columnwidth]{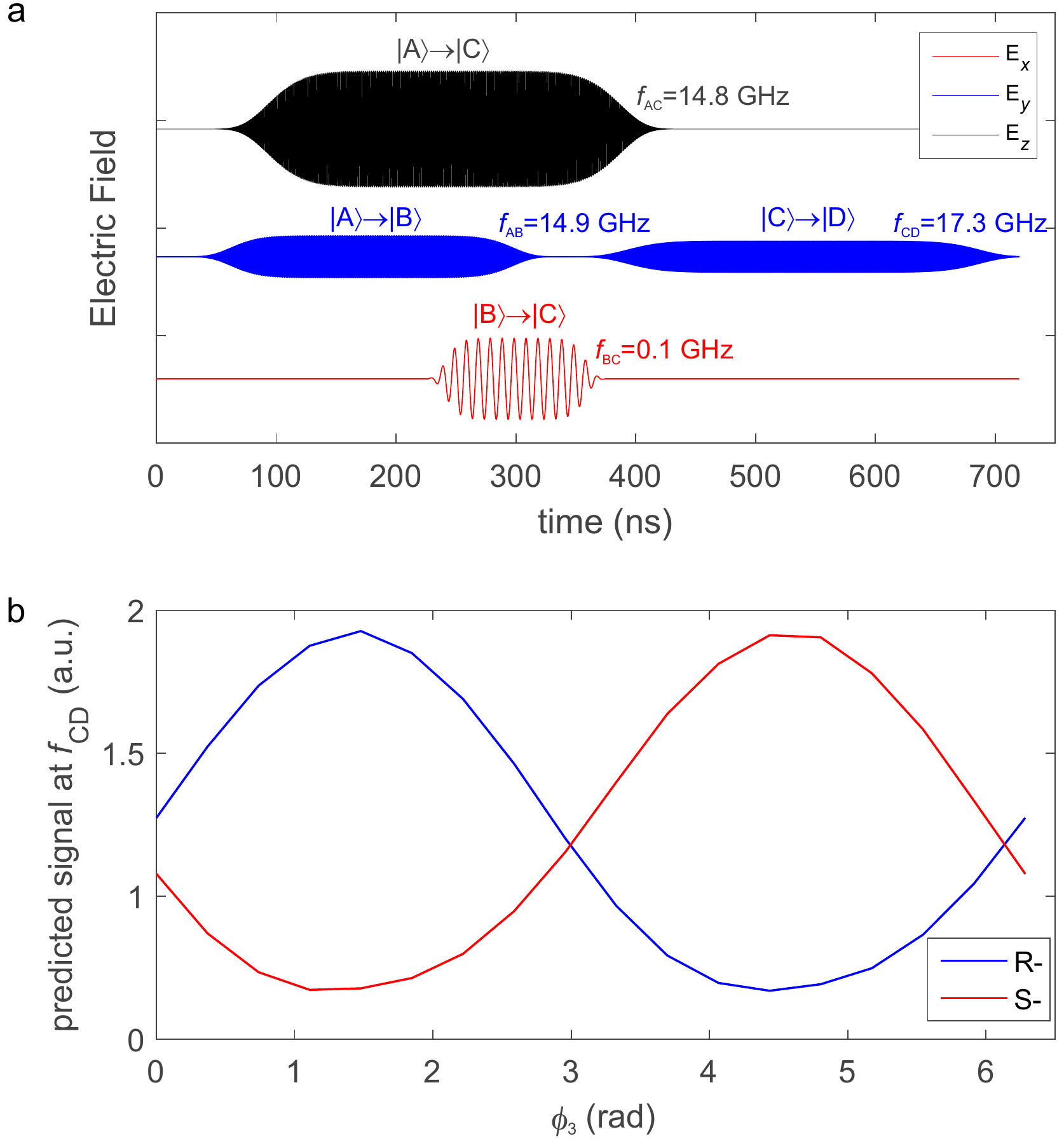}
\caption{Applied microwave pulses and predicted state-specific enantiomeric enrichment signal. a) The applied four-pulse sequence used for the state transfer. The first three pulses selectively promote either $R$ or $S$ enantiomers to the excited state $|C\rangle$. The fourth pulse ($|C\rangle \rightarrow |D\rangle$) is used for probing the population in the target state. b) The predicted amplitude of the signal at $f_{CD}$ as a function of the phase $\phi_3$ of the drive pulse $|B\rangle \rightarrow |C\rangle$.  Similar sinusoidal behavior is predicted as a function of $\phi_1$ and $\phi_2$. For the simulation, field-free Hamiltonian and transition matrices are calculated~\cite{Pgopher}, and the time-dependent Schr\"{o}dinger equation is numerically integrated.  
}\label{fig:phasepredict}
\end{figure}

A full simulation of enantiomer-specific state transfer for 1,2-propanediol is shown in figure~\ref{fig:phasepredict}. This simulation does not make the approximation of the weak pulse limit. In figure \ref{fig:phasepredict}b, the predicted signal amplitude at $f_{CD} = 17343\,\mathrm{MHz}$ is shown for $R$ and $S$ molecules as a function of the drive phase $\phi_{3}$. The predicted sinusoidal dependence of the signal amplitude on $\phi_1$, $\phi_2$, and $\phi_3$ provides an unambiguous fingerprint that the molecules are in fact traversing the interfering trajectories $|A\rangle \rightarrow |C\rangle$ and $|A\rangle \rightarrow |B\rangle \rightarrow |C\rangle$. 

\begin{figure}[ht]
\includegraphics[width=0.5\columnwidth]{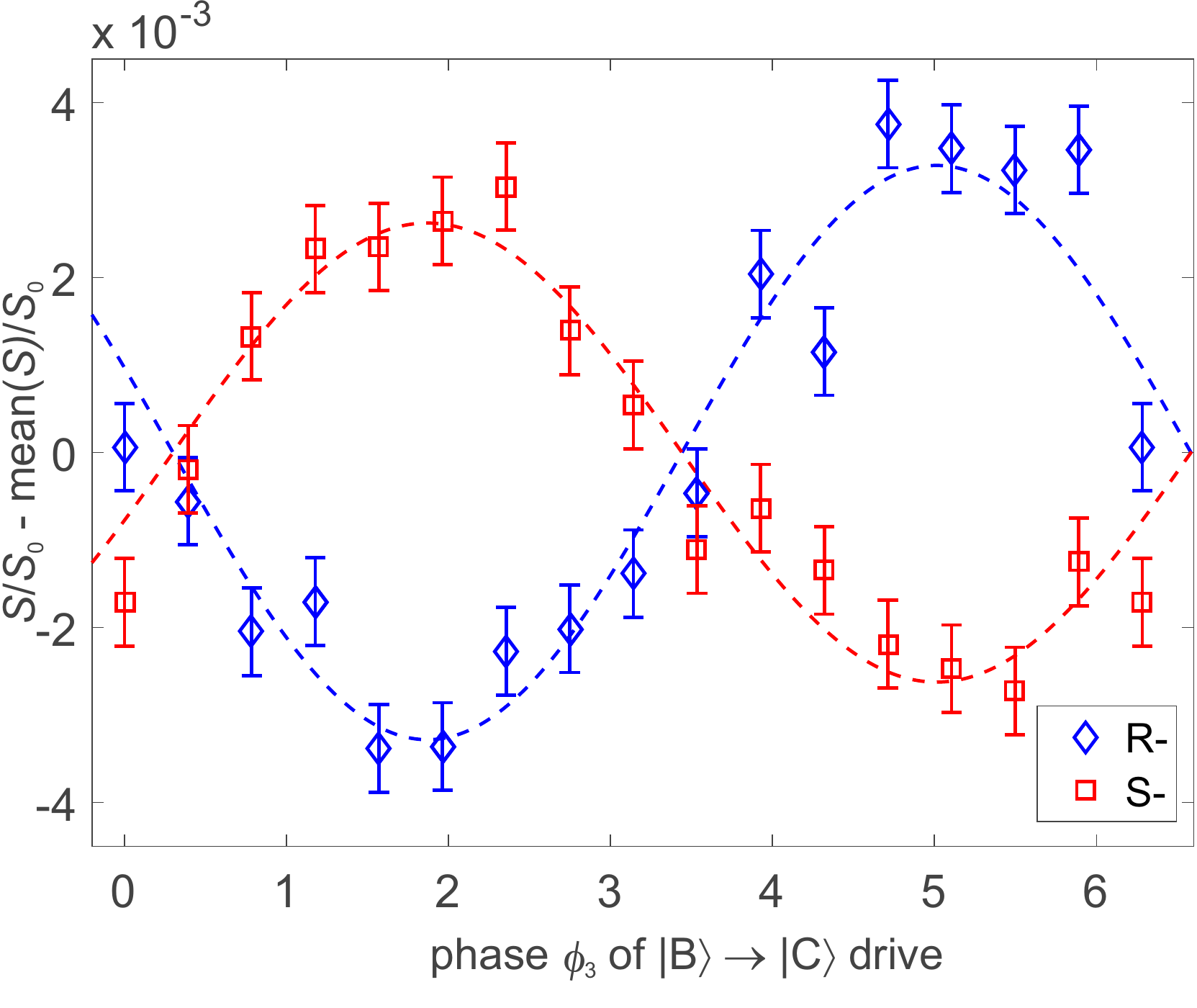}  
\caption{Enantiomer-specific enrichment of the rotational state $|C\rangle$ of chiral 1,2-propanediol. Variations in the amplitude $S$ of the FID signal of the $|C\rangle \rightarrow |D\rangle$ transition reflect the changing population in $|C\rangle$. The red and blue data show the signal from enantiopure $S$ and $R$ 1,2-propanediol, respectively. The data show the change in the population of state $|C\rangle$, represented by the quantity $S/S_0-\mathrm{mean}(S)/S_0$, as a function of the phase $\phi_3$ of the $|B\rangle \rightarrow |C\rangle$ drive. Here, $S_0$ is the signal amplitude for the transition $|C\rangle\rightarrow|D\rangle$ with all of the drive pulses off and the $|C\rangle \rightarrow |D\rangle$ probe pulse on, and $\mathrm{mean}(S)= \frac{1}{N}\sum_i (S_i)$ is the mean measurement $S$ over all phases $\phi_3$ with $N$ being the number of measurements and $i$ being the index of the measurement. For further information see the Supplemental Material. The enantiomer-specific phase dependence is clearly visible. The red and blue lines show sinusoidal fits to the data of $S$ and $R$ molecules, with root mean squared errors of $7.0\times 10^{-4}$ and $7.3\times 10^{-4}$, respectively. With an initially racemic sample, the maximum achieved state-specific enrichment is proportional to the peak-to-peak amplitude of the sinusoidal fit and amounts to $\epsilon = 0.54\pm 0.05\,\%$ (Supplemental Material).} \label{fig:enrichfigure}
\end{figure}

The enantiomer-specific state transfer is demonstrated with enantiopure samples of $R$ and $S$ 1,2-propanediol. The experimental data resulting from the above described pulse sequence are shown in figure~\ref{fig:enrichfigure}. The whole experimental cycle, including all microwave pulses, needs to be completed within a time not much longer than $\tau_c \approx 6\,\mathrm{\mu s}$, because every collision resets the rotational coherence. Because of microwave power limitations, we chose the longest possible pulse lengths below this limit. The experimental pulse timing is shown in the Supplemental Material. While the phase $\phi_3$ of the $|B\rangle \rightarrow |C\rangle$ drive is varied, the population in $|C\rangle$ is recorded by measuring the amplitude $S$ of the FID signal of the $|C\rangle \rightarrow |D\rangle$ transition. The experimental data are shown in figure~\ref{fig:enrichfigure}. As predicted, we observe a sinusoidal dependence on the drive phase. 

Even given perfect pulse fidelity, two factors limit the enrichment in state $|C\rangle$. First, the initial population in state $|A\rangle$ is distributed equally over several $m_J$ states, which are connected to states $|B\rangle$ and $|C\rangle$ by different matrix elements, e.g. $\langle 2210|E_z|2200\rangle \neq \langle 2211|E_z|2201\rangle $ in the $|JK_aK_cm_J\rangle$ notation. The optimal pulse amplitudes to achieve efficient population transfer depend on $m_J$; in the case of the population being distributed over several $m_J$ states, the pulses cannot simultaneously fulfill the desired optimal pulse conditions for all $m_J$ states. This limits the predicted transfer fidelity of $|A\rangle \rightarrow |C\rangle$. In addition, our method \emph{exchanges} the population of $R$ ($S$) molecules in states $|A\rangle \rightleftharpoons |C\rangle$. Therefore, at nonzero temperature, due to the thermal distribution of occupied states, there will be thermally excited $S$ ($R$) molecules remaining in state $|C\rangle$ throughout the exchange process, even if the experimental settings are optimized for enantiomer-specific state enrichment for $R$ ($S$) molecules. The maximum enrichment $\epsilon =2 \frac{P_S - P_R}{P_S+P_R}$ for such a process for the present cell temperature of  $T = 10\,\mathrm{K}$, is $\epsilon \approx 5\,\%$. Here, $P_R$ and $P_S$ are the fractions of $R$ and $S$ molecules populating state $|C\rangle$, with $P_S + P_R = 1$. We measure clear state-specific enantiomeric enrichment and depopulation in the target state for opposite enantiomers (figure~\ref{fig:enrichfigure}). The demonstrated state-specific enantiomeric enrichment corresponds to $\epsilon = 0.54 \pm 0.05\,\%$ when starting from a racemic (equal amounts of $S$ and $R$ enantiomers) sample. This modest enrichment could be significantly increased by increasing the power of the applied microwave pulses (the current experiments were limited by the available power amplifiers) and/or by decreasing the cell temperature. Lowering the temperature to $T \lesssim 1K$, which is routinely achieved in supersonic beams and in several buffer gas cooling experiments, would increase $\epsilon$ by a factor of $\sim 10$. Choosing states with higher transition frequencies, or adding additional pulses to exchange the initial unwanted population of molecules in $|C\rangle$ with higher, less populated states would further increase $\epsilon$. In combination, these measures would significantly increase the enrichment by a factor of $\sim 50$, to $\epsilon \approx 30\,\%$.

The sensitivity of this method to very slight enantiomeric excess could be dramatically improved by replacing the final readout pulse, $|C\rangle \rightarrow |D\rangle$, with a more sensitive, state-selective detection method. The probed transition could be any rotational, vibrational, or electronic transition. This could be realized with LIF, resonance-enhanced multiphoton dissociation, or REMPI, all of which provide sensitive measurements of the population in a given rovibrational state, but are in general not sensitive to the phase between two states. The readout could also be achieved using infrared spectroscopy, which has been recently demonstrated in a similar buffer gas environment~\cite{Spaun2016}.

Our method of state-specific enantiomeric enrichment is an attractive resource for high resolution PV searches~\cite{Darquie2010,Schwerdtfeger2010,Quack2008}.
Small frequency differences between $S$ and $R$ molecules resulting from the parity violating nuclear weak force are predicted, but so far have not been observed. The magnitude of the predicted frequency shift $\Delta f$ varies dramatically with molecule, but the fractional shift $\Delta f / f$ is typically less than $10^{-14}$~\cite{Quack2008,Saleh2013}. Measurement of these frequency differences will require spectroscopy of high precision, similar to that of searches for the electron electric dipole moment~\cite{Baron2014}. The method demonstrated here allows for such a measurement on enantiomers from within a racemic mixture, avoiding the requirement for enantiopure samples. In many molecules, the preparation of chirally pure samples is intrinsically difficult because of rapid racemization, either via thermal excitations in room temperature solutions, or via tunneling between enantiomers. To our knowledge, the method demonstrated here is the only demonstrated method that could study enantiomers that racemize in a time $5\,\mathrm{\mu s} < \tau < 1\,\mathrm{s}$. In addition, the enantiomers can be selected simply by changing the phase $\phi_1$, $\phi_2$, or $\phi_3$, allowing for rapid alternation of enantiomer with remarkably few opportunities for the introduction of systematic errors, often the limiting factor in such precision experiments. 

In conclusion, we have demonstrated a new, widely applicable, nonchemical method to create state-selective enantiomeric excess of chiral molecules. This could provide an important new tool for the search of parity violating terms in the spectra of chiral molecules, or other effects of chirality. 

\begin{acknowledgments}
This work has been supported by the US National Science Foundation (NSF CHE-1506868). S.E. acknowledges funding through the Schr{\"o}dinger Fellowship of the Austrian Science Fund (FWF): J3796-N36.
\end{acknowledgments}

\section{Supplemental Material to\\``Enantiomer-Specific State Transfer of Chiral Molecules"}
\begin{center}
Sandra Eibenberger, John Doyle, and David Patterson\\
\textit{Harvard University, Department of Physics, 17 Oxford Street, Cambridge, MA 02138 USA}
\end{center}

\subsection{Application to parity violation experiments}
Our method of state-specific enantiomeric enrichment is an attractive resource for high resolution parity violation (PV) searches~\cite{Darquie2010,Schwerdtfeger2010,Quack2008}.

\begin{figure}[ht]
\includegraphics[width=\columnwidth]{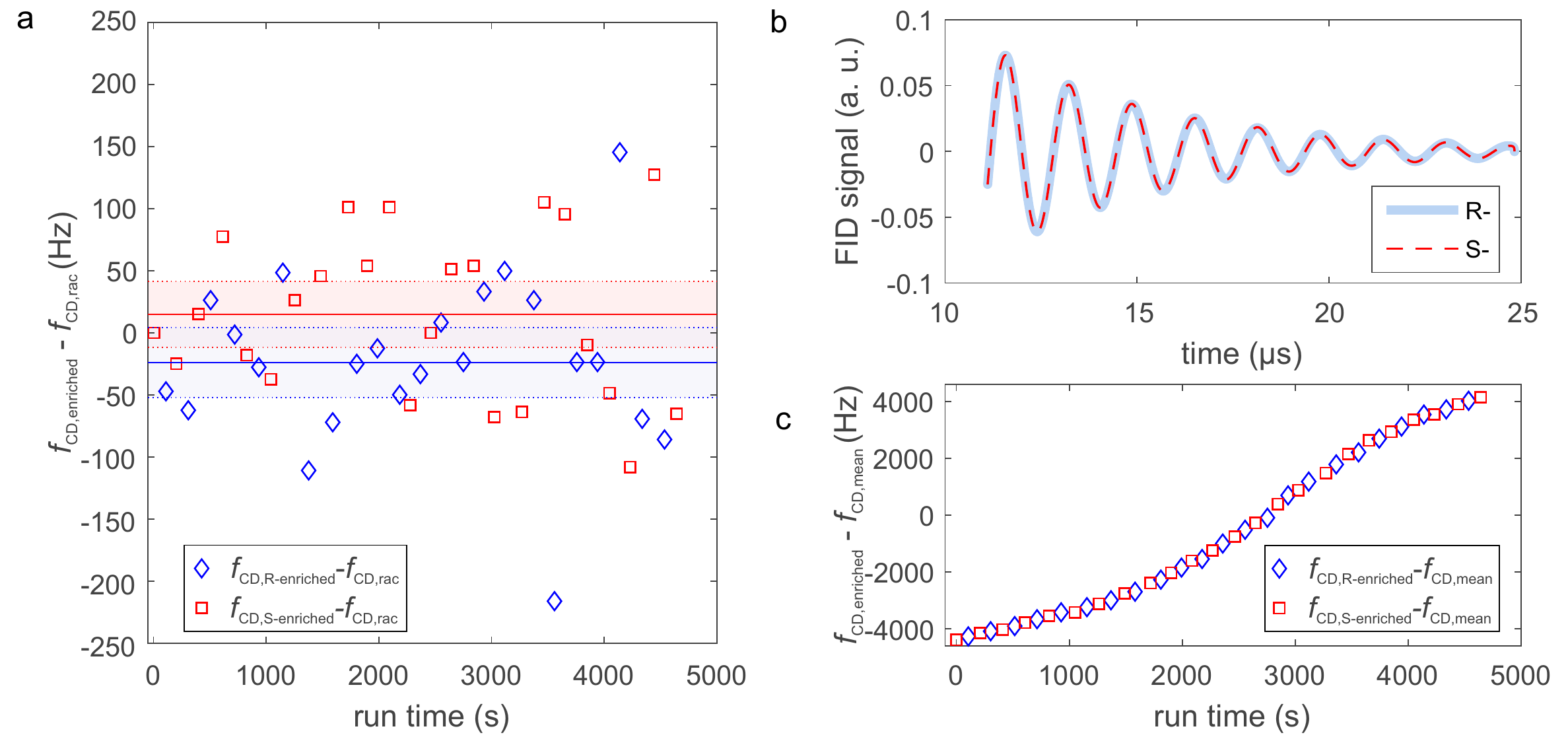}  
\caption{a) Consecutive differential measurements of $f_{\mathrm{CD,S\text{-}enriched}}-f_{\mathrm{CD,rac}}$ and $f_{\mathrm{CD,R\text{-}enriched}}-f_{\mathrm{CD,rac}}$. Here, $f_{\mathrm{CD,rac}}$ is the measured frequency of the transition $|C\rangle \rightarrow |D\rangle$ of the racemic sample with the enantiomer-specific state transfer pulses all off, and $f_{\mathrm{CD,S\text{-}enriched}}$ and $f_{\mathrm{CD,R\text{-}enriched}}$ are the measured frequencies of the transition $|C\rangle \rightarrow |D\rangle$ with the enantiomer-specific state transfer pulses on and the phase $\phi_3$ optimized for $S$ and $R$ molecules within the racemic sample, respectively. The red ($S$-enriched) and blue ($R$-enriched) shaded areas represent the $95\,\%$ confidence intervals for the standard error of the mean measured frequency differences. The measured frequency differences amount to $f_{\mathrm{CD,S\text{-}enriched}}-f_{\mathrm{CD,rac}}=15 \pm 27\,\mathrm{Hz}$ and $f_{\mathrm{CD,R\text{-}enriched}}-f_{\mathrm{CD,rac}}=-24\pm 28\,\mathrm{Hz}$. The stated errors correspond to the  $95\,\%$ confidence interval for the standard errors of the mean values. b) Free induction decay (FID) signals, mixed down in frequency for visibility, for the state transfer optimized for state-specific enrichment of $S$ (red) and $R$ (blue) molecules. As expected, there is no significant difference measured. c) Consecutive measurements of the frequency $f_{\mathrm{CD,S\text{-}enriched}}-f_{\mathrm{CD,mean}}$ and $f_{\mathrm{CD,R\text{-}enriched}}-f_{\mathrm{CD,mean}}$, with $f_{\mathrm{CD,mean}}$ being the mean measured transition frequency, show a drift in the order of $7\,\mathrm{kHz}$ over approximately one hour of measurement time.} \label{fig:parityfigure}
\end{figure}

\begin{figure}[ht]
\includegraphics[width=\columnwidth]{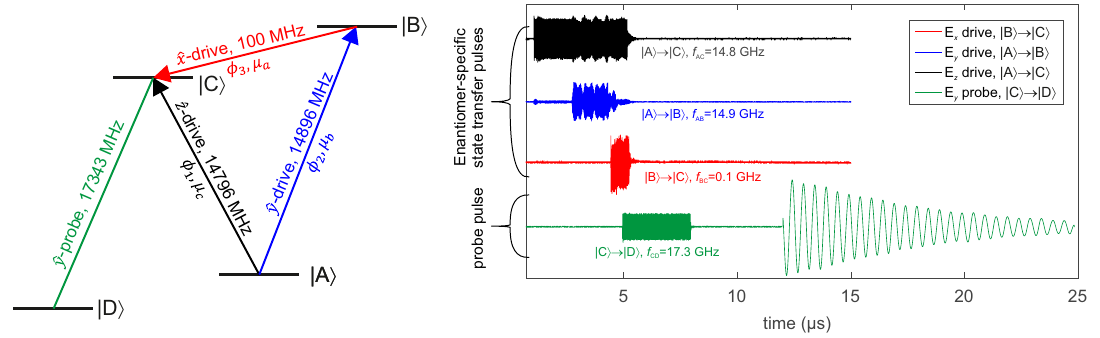} 
\caption{Left: Level diagram of the rotational states of 1,2-propanediol we employ for enantiomer-specific enrichment of a chosen rotational state. The relevant states are labeled $|A\rangle$, $|B\rangle$, $|C\rangle$, and $|D\rangle$, corresponding to $|A\rangle = |211\rangle$, $|B\rangle = |220\rangle$, $|C\rangle = |221\rangle$, and $|D\rangle = |212\rangle$, in the $|JK_aK_c\rangle$ notation. Right: The experimental pulse sequence. $R$ or $S$ enantiomers are selectively swapped between the states $|A\rangle$ and $|C\rangle$. In the lowest trace, a typical free induction decay, recorded at $17343\,\mathrm{MHz}$ ($|C\rangle \rightarrow |D\rangle$), and mixed down for visibility, is shown. Variations in the amplitude $S$ of this signal reflect the changing population in $|C\rangle$. Starting at time $t_0=0$, there is applied a $4.4\,\mathrm{\mu s}$ long, $\sim 0.3\,\mathrm{V/cm}$, pulse at $f_{AC} = 14796\,\mathrm{MHz}$ ($|A\rangle \rightarrow |C\rangle$); at $t=1.7\,\mathrm{\mu s}$, a $1.9\,\mathrm{\mu s}$ long, $\sim 0.4\,\mathrm{V/cm}$, pulse at $f_{AB} = 14896\,\mathrm{MHz}$ ($|A\rangle \rightarrow |B\rangle$); at $t=3.6\,\mathrm{\mu s}$, a $0.8\,\mathrm{\mu s}$ long, $\sim 1.1\,\mathrm{V/cm}$, pulse at $f_{BC} = 100.5\,\mathrm{MHz}$ ($|B\rangle \rightarrow |C\rangle$). The phase $\phi_3$ of this pulse ($|B\rangle \rightarrow |C\rangle$) is varied to selectively transfer $R$ or $S$ 1,2-propanediol into state $|C\rangle$. At $t=4.4\,\mathrm{\mu s}$, a probe pulse at $f_{CD} = 17343\,\mathrm{MHz}$ is applied.} \label{fig:exptiming}
\end{figure}

In order to illuminate this method as a tool in a search for PV in molecular spectra, we perform enantiomer-specific spectroscopy on the $|C\rangle \rightarrow |D\rangle$ transition of racemic 1,2-propanediol. We alternatingly produce enantiomeric excess in state $|C\rangle$ for $S$ and $R$ molecules, using our method of state-specific enantiomeric enrichment. The transition $|C\rangle \rightarrow |D\rangle$ is driven and the FID signal for $S$ enriched and $R$ enriched 1,2-propanediol, selected from a racemic mixture, is alternatingly recorded and compared to the signal produced with the enantiomer-specific state transfer pulses off. The resulting data are shown in figure~\ref{fig:parityfigure}. The experimental pulse sequence for the enantiomer-specific state transfer is shown in figure~\ref{fig:exptiming}. This type of enantiomer-specific spectroscopy constitutes an essential ingredient in a future PV experiment. We observe a frequency drift in the order of $7\,\mathrm{kHz}$ over approximately one hour of measurement time, as shown in figure~\ref{fig:parityfigure}c. A likely cause for this frequency drift is the changing environment in the cryogenic buffer gas cell, with drifting static electric fields introducing uncontrolled Stark shifts. A rigorous search for PV in molecular spectra would need to include a careful analysis of systematic effects and determination of systematic errors, which is not included here. In addition, we did not blind our experiment as is appropriate in rigorous precision measurement. We note, however, that the rapid switching between $R$ and $S$ enantiomers - based on only a phase shift - is strongly expected to lead to substantial immunity from systematic errors. 
Our measurements yield values $f_{\mathrm{CD,S\text{-}enriched}}-f_{\mathrm{CD,rac}}=15 \pm 27\,\mathrm{Hz}$ and $f_{\mathrm{CD,R\text{-}enriched}}-f_{\mathrm{CD,rac}}=-24\pm 28\,\mathrm{Hz}$, resulting in $\Delta f_{\mathrm{measured}}=f_{\mathrm{CD,S\text{-}enriched}} - f_{\mathrm{CD,R\text{-}enriched}}=39\pm 39\,\mathrm{Hz}$, with the stated errors corresponding to the $95\,\%$ confidence intervals for the standard errors of the mean values. 
In the limit of $\Delta f = f_{\mathrm{CD,S}} - f_{\mathrm{CD,R}} \ll R$, and $\epsilon \ll 1$, with $R$ being the resolution, a nonzero $\Delta f$ would manifest as a measured frequency shift $\Delta f_{\mathrm{measured}} = \frac{\epsilon}{2} \Delta f$. Here, $f_{\mathrm{CD,S}}$ and $f_{\mathrm{CD,R}}$ are the frequencies of the $|C\rangle \rightarrow |D\rangle$ transition of enantiopure $S$ and $R$ 1,2-propanediol respectively. For our modest enrichment of $\epsilon \approx 0.54\,\%$, our measurement translates to $\Delta f = 14\pm 14\,\mathrm{kHz}$, with the stated error being statistical.

\subsection{Constants, pulse lengths, and field amplitudes of the simulation}
The rotational constants $A$, $B$, $C$, and electric dipole moments $\mu_a$, $\mu_b$, $\mu_c$, of the lowest energy conformer of 1,2-propanediol are used as determined in~\cite{Lovas2009}: $A = 8572.0553\,\mathrm{MHz}$, $B = 3640.1063\,\mathrm{MHz}$, $C = 2790.9666\,\mathrm{MHz}$, $\mu_a = 1.201\,\mathrm{D}$, $\mu_b = 1.916\,\mathrm{D}$, and $\mu_c = 0.365\,\mathrm{D}$. The simulated pulse lengths and electric field amplitudes are: $150\,\mathrm{ns}$, $10\,\mathrm{V/cm}$ ($|A\rangle \rightarrow |C\rangle$); $120\,\mathrm{ns}$, $6\,\mathrm{V/cm}$ ($|A\rangle \rightarrow |B\rangle$); $60\,\mathrm{ns}$, $15\,\mathrm{V/cm}$ ($|B\rangle \rightarrow |C\rangle$)

\subsection{Experimental pulse sequence}

\subsection{Determination of the achieved state-specific enantiomeric enrichment}
Before all applied pulses, the state $|C\rangle$ is occupied by $N_S$ $S$ molecules and $N_R$ $R$ molecules. For enantiopure samples, $N_S=0$ or $N_R=0$. For racemic samples $N_S=N_R$.
The final populations in state $|C\rangle$ are
\begin{equation}
P_S = N_S(1+T_S) \text{ and  } P_R=N_R(1+T_R).\label{eq:Pdef}
\end{equation}
Here, $T_S$ and $T_R$ designate the transition coefficients to any other state.
$T_R>0$ means that $R$ molecules are brought into $|C\rangle$; $T_R<0$ means that $R$ molecules are brought out of state $|C\rangle$.
Our measurable is the signal $S \propto P_S+P_R$. We define $k$ by 
\begin{equation}
S=k(P_S+P_R).\label{eq:S}
\end{equation}
With the enantiomer-specific state transfer pulses off and with the probe pulse on (the case of $T_S=T_R=0$, $P_S=N_S$ and $P_R=N_R$) we measure 
\begin{equation}
S_0=k(N_S+N_R).\label{eq:S0}
\end{equation}
With all enantiomer-specific state transfer pulses on and probe pulse on, but with a phase $\phi_3$ chosen such that there is no enantiomer-specific state transfer (the case of $T_S=T_R=T$) we measure $S_{\Delta}=k(P_S+P_R)=k(N_S(1+T)+N_R(1+T))$, leading to
\begin{equation}
\frac{S_{\Delta}}{S_0}=1+T.
\end{equation}
In essence, $S_{\Delta}$ is equivalent to $\mathrm{mean}(S)$.
With the enantiomer-specific state transfer pulses being on, with a $\phi_3^*$ chosen such that the measurable $S/S_0-\mathrm{mean}(S/S_0)$ is maximum, $y=y_{\mathrm{max}}$, we have $T_S=T+\alpha$ and $T_R=T-\alpha$. Now, $\alpha$ can be related to $y_{\mathrm{max}}$.
Using equation~\ref{eq:Pdef} results in $P_S=N_S(1+T_S)=N_S(1+T+\alpha)$ and $P_R=N_R(1+T_R)=N_R(1+T-\alpha)$.
By using equation~\ref{eq:S}, $S=k(N_S(1+T+\alpha)+N_R(1+T-\alpha))$.
After dividing this by $S_0$ and using equation~\ref{eq:S0},
\begin{equation}
\frac{S}{S_0}=\frac{(N_S+N_R)(1+T)}{(N_S+N_R)}+\frac{\alpha(N_S-N_R)}{(N_S+N_R)},
\end{equation}
resulting in 
\begin{equation}
y=\frac{S}{S_0}-\mathrm{mean}\left(\frac{S}{S_0}\right)=\frac{\alpha(N_S-N_R)}{(N_S+N_R)}.
\end{equation}
Assuming $N_S=1$, $N_R=0$, and $y=y_{\mathrm{max}}$, we set $y_{\mathrm{max}}=\alpha$.
When operating at $\phi_3^*$ using a racemic sample of molecules the final populations are
\begin{equation}
P_S=N_S(1+T+y_{\mathrm{max}}) \text{ and  } P_R=N_R(1+T-y_{\mathrm{max}})
\end{equation}
Defining the enrichment as
\begin{equation}
\epsilon=2\frac{(P_S-P_R)}{(P_S+P_R)},
\end{equation}
and with $N_S=N_R$ for a racemic mixture, this results in 
\begin{equation}
\epsilon=2\frac{N_S(1+T+y_{\mathrm{max}}-1-T+y_{\mathrm{max}})}{2N_S(1+T)}=2\frac{y_{\mathrm{max}}}{1+T}.
\end{equation}
In our measurements, $1+T=\mathrm{mean}\left(S/S_0\right)\approx 1.17$, and, resulting from the amplitude of a sinusoidal fit to the data, $y_{\mathrm{max}}=0.0032$, resulting in a maximum achieved enrichment of $\epsilon=2y_{\mathrm{max}}/1.17=0.54\%$.

%


\end{document}